# Evaluation of Lightweight Block Ciphers in Hardware Implementation: A Comprehensive Survey

Jaber Hossein Zadeh
Data and Communication Security Laboratory (DCSL)
Faculty of Engineering, Ferdowsi University of Mashhad
Mashhad, Iran
Jaber_hosseinzadeh@stu-mail.um.ac.ir

Abbas Ghaemi Bafghi
Data and Communication Security Laboratory (DCSL)
Department of Computer Engineering, Faculty of Engineering ,Ferdowsi University of Mashhad
Mashhad, Iran
Ghaemib@um.ac.ir

*Abstract*— **The conventional cryptography solutions are ill-suited to strict memory, size and power limitations of resource-constrained devices, so lightweight cryptography solutions have been specifically developed for this type of applications. In this domain of cryptography, the term lightweight never refers to inadequately low security, but rather to establishing the best balance to maintain sufficient security. This paper presents the first comprehensive survey evaluation of lightweight block ciphers in terms of their speed, cost, performance, and balanced efficiency in hardware implementation, and facilitates the comparison of studied ciphers in these respects. The cost of lightweight block ciphers is evaluated with the metric of Gate Equivalent (Fig.1), their speed with the metric of clock-cycle-per-block (Fig.2), their performance with the metric of throughput (Fig.3) and their balanced efficiency with the metric of Figure of Merit (Fig.4). The results of these evaluations show that SIMON, SPECK, and Piccolo are the best lightweight block ciphers in hardware implementation.**(*Abstract*)

### *Introduction*

Lightweight cryptography has been developed specifically for low-cost resource-constrained devices, as its design allows it work with limited hardware. Devices used in wireless sensor networks, RFID tags, and Internet of things (IoT) are mostly characterized by low computing power, limited batteries, low memory, low power consumption and low operating frequency range [1, 11, 2, 24, 25]. These devices are often employed in poorly accessible and sometimes critical environments (e.g. in military applications) and work with limited batteries and an insecure communication channel, and all these factors highlight their need to robust cryptographic solutions [4, 5, 11, 24, 25]. On the other hand, the high computation and energy requirements of common cryptography methods such as AES, RSA emphasize the focus on lightweight solutions. So the growing use and development of resource-constrained devices such as smart phones, smart cards, etc. and the rising importance of security as their core principle has led to increased interest to lightweight cryptography [1, 2, 5, 25]. The lightweight symmetric ciphers can be categorized into two classes: Block-based and stream-based. The following is a brief introduction to some of the lightweight block ciphers available in the literature.

**SEA**: This cipher was designed in 2006 by Standaert et al. The design of this cipher is based on low memory requirements, minimal code size, and limited instruction set, plus flexibility, which is an unusual design criterion for ciphers. This cipher is based on Feistel structure and it can work with different text, key, and word sizes. This cipher is denoted by $SEA_{n,b}$, where $n$ is the plaintext size and key size, and $b$ is the processor (or word) size. Due to its simplicity constraints, this cipher employs a limited number of basic operations, such as bitwise XOR, substitution box S, word (left) rotation, inverse word rotation, bit rotation, and modular addition [27].

**HIGHT:** This cipher was developed by Deukjo Hong et al. in 2006. It uses a 64-bit block size and a 128-bit key size. Its basic structure is 32-round type-2 generalized Feistel Network (GFN-2). The encryption processing of this cipher starts with initial conversion of the block, continues with a 32-round iterative function, and ends with final transform of the output of round function. The mentioned round function employs two functions $F_0$ and $F_1$ plus XOR and addition operations. Functions $F_0$ and $F_1$ are based on simple XOR and shift operations [20].

**Hummingbird:** This cipher was introduced in 2010 by Daniel Engels et al. It has a hybrid structure composed of block- and stream-based designs. It employs a 16-bit block size, a 256-bit key size and an 80-bit internal state. The size of the key and the internal state of Hummingbird provides an adequate level of security for many embedded applications. The overall structure of the Hummingbird encryption algorithm uses four 16-bit block ciphers $E_{k1}$,$E_{k2}$,$E_{k3}$,$E_{k4}$, plus 16-bit internal state registers, and a 16-stage LFSR. Each block cipher has a 16-bit substitution-permutation structure and a 64-bit key size. In the SPN structure, the block-based part of the cipher uses the XOR operation for Key Addition, four 4-bit different S-boxes for substitution layer, and a XOR-included linear transform [3].

**PRESENT**: This cipher, which was developed in 2007 by A. Bogdanov et al, is based on a substitution-permutation structure, 64-bit blocks, and 80-bit keys. Addition part of round key consists of simple XOR operation. The substitution layer is composed of sixteen 4-bit S-boxes and the permutation layer consists of bitwise permutation. This algorithm runs a 31-round iteration to return a ciphertext. In 2012, this cipher was approved by International Organization for Standardization (ISO / IEC 29192-2) as a standard lightweight block cipher [15].

**PRINTcipher:** In 2010, Lars Knudsen et al. designed this cipher specifically for IC-printing. The aim of their design was to ensure memory persistence. This design has two versions, 48-bit and 96-bit. The 48-bit version uses a 48-bit secret key. This cipher uses b-bit blocks ($b \in \{48, 96\}$) and an effective key length of (5/3*b)-bits, and its structure is based on b-round substitution-permutation network. For instance, the 48-bit version of this cipher uses 48-bit blocks and 80-bit key and enjoys a 48-round structure. The encryption process of this cipher starts with a 48-bit mapping on the input; cipher then applies one round of XOR on the 6 least significant bits, and subjects the output to key-dependent permutation and then to substitution layer. Substitution layer of this cipher consists of sixteen 3-bit S-boxes. The output of this layer is the output of one round. As mentioned earlier, the PRINTcipher-48 employs 48 rounds, which means 48 iteration of the described process [8].

**KATAN&KTANTAN:** Christophe De Canniere et al. developed this family of ciphers in 2009. Both versions utilize 32, 48 and 64-bit block size, and share 80-bit key and security level. KTANTAN is the compact version of the cipher, where the key is burnt into the device and cannot be changed. In these ciphers, the plaintext is loaded into two registers. In each round, cipher selects several bits of registers, subjects them to Boolean functions and then loads the output into the least significant bits of the shifted registers. This cipher needs 254 rounds of iteration to ensure sufficient mixing [12].

**mCrypton**: This cipher, which was developed in 2005 by Chae Hoon Lim et al., uses 64-bit blocks and 64,96, or 128-bit key sizes. The main objective of this cipher is to optimize the efficiency for resource-constrained applications. mCrypton processes the 8-bit data blocks 4 expressed as 4 by 4 nibble array. Each round of transformation consists of 4 operations: nibble-wise substitution, column-wise bit permutation, column-to-row transposition, and key addition. The encryption process of this cipher consists of 12 iterations of round transformation [13].

**KLEIN**: This cipher was designed by Zheng Gong et al in 2011. The basic structure of this cipher is based on substitution-permutation network (SPN), and it has been designed with round counts of 12, 16, and 20 for 64, 80, and 98 bit variations. The cipher's input and output are in the form of one-dimensional array of bytes. In this cipher, operations are optimized with byte-oriented algorithms. Like many other SPN-based ciphers, the stage of Add-Round-key is implemented via simple XOR operation. The substitution stage uses 16 similar involutive S-boxes; this involution property means $S(x)=y$, $S(y)=x$ and $S(S(x))=x$.

The advantage of using an involutive s-box is the reduction of extra cost of inverse implementation which leads to efficient serialization [7].

**TWINE**: This cipher was developed in 2013 by T. Suzaki et al. It uses 64-bit block size and 80 or 128 bit key size. The design of this cipher is geared toward desirable hardware and software performance on different types of central processors. This design is based on type-2 generalized Feistel Network (GFN-2) with sixteen nibble blocks. This cipher partitions the 64-bit block to sixteen $X_i$ , and in line with GFN-2 structure, uses 8 simple F functions. The X's having an even subscript proceed to the next stage as they are, but they are inserted into the positions set by 4-bit-wise permutation. Cipher also imports the X's with even subscripts into F function and XORs them with the X's having an odd subscript. Here, permutation employs 4-bit words and forms the linear part of the cipher [18].

**SIMON**: In 2013, Ray Beaulieu et al. developed this family of ciphers with different block and key sizes. SIMON2n uses n-bit words (in this case block size is 2n), where n can be 16, 24, 32, 48, or 64- bit. This SIMON2n / mn uses m-word (mn-bit) key. For instance SIMON64/128 will employ 64-bit blocks and 128-bit keys. All SIMON ciphers use the same Feistel rule. The algorithm of these ciphers is engineered to be easily serialized at different levels of extremely small hardware, but not at the expense of software performance [22].

**SPECK**: This cipher was specifically designed to provide optimized hardware and software performance on microcontrollers. Nomenclature of SPECK is similar to that explained for SIMON. For instance, SPECK96/144 will use 96-bit block and 144-bit key size. This cipher utilizes bitwise XOR, modular addition $2^n$, left circular shift $S^j$ by j bits, and right circular shift $S^{-j}$ by j bits [22].

**PRINCE**: This cipher was designed in 2012 by Julia Borghoff et al. PRINCE uses 64-bit block size and 128-bit key size and is based on FX structure. The cipher employs a Key Whitening component to spread the effect of key throughout the plaintext and prevent key-based attacks. Between the key whitening parts is the 12-round PRINCE core. This core consists of simple XOR, addition of round constant, plus substitution and Matrix-M operations. This design uses similar 4-bit S-boxes, and twelve 64-bit round constants [16].

**PRIDE**: In 2012, Martin R. Albrecht et al. developed the PRIDE cipher, which like PPRINCE, is based on FX structure. This cipher uses 64-bit block size and 128-bit key size. This cipher extracts the first whitening key k from the first half the key k and uses the other half to obtain the second whitening key $k_1$ . To ensure effective bit-sliced implementation, it uses a bit permutation at the start and the end of process. The encryption process of this cipher starts with an initial bit permutation on plaintext. Cipher then subjects the results to an XOR with the first whitening key. It then applies 19 identical rounds of iteration on the output. The $20^{th}$ round, which is applied on the output of round 19, is slightly different than the others. Cipher then XORs the results with the second whitening key and then applies the secondary bit permutation on the result. The output of this process will be the ciphertext c. The round function R, which is applied on the first 19 rounds, is a classical substitution-permutation network. In this function, the key Addition stage is implemented by a XOR. The substitution layer consists of sixteen 4-bit S-boxes. The linear parts of this function include the first bit permutation, the L function, and the second bit permutation. The $20^{th}$ round includes only the substitution layer [19].

**Hummingbird2**: Daniel Engels et al. developed the HB2 in 2012. This cipher uses a 128-bit secret key and a 64-bit initialization vector. The main advantage of this cipher is its ability to produce authentication tags for each selectively processed message. This cipher has a 128-bit internal state which is initialized by a 64-bit array. HB2 is a hybrid construct composed of block and stream ciphers and, like HB, works with 16-bit block size. So its operations have been designed for 16-bit words. This cipher uses a nonlinear F function, which has been defined by a linear operation on 4 different nonlinear S-boxes. This means that the input of linear function is the output of non-linear function (S-box) [28].

**LBLOCK**: This cipher was introduced in 2011 by Wenling Wu et al. It works with 64-bit block size and 80-bit key size, and is based on 32-round Feistel structure. The security of Feistel structure is associated with the round function F. The round function of this cipher is composed of two parts, S and P, which establish the basic Shannon principles. The substitution layer S is responsible for clutter operation and the permutation layer P diffuses the Shannon principles. The substitution layer has eight parallel 4-bit S-boxes, and the permutation layer consists of eight 4-bit permutations, i.e. the basic element of this permutation works with 4 bits. It should be mentioned that this cipher uses 8 different S-boxes [9].

**MIBS**: Maryam Izadiet al. designed the MIBS cipher in 2009. This Feistel-based 32-round cipher uses 64-bit blocks and 46 and 80-bit keys. The round function of this cipher consists of 8 identical S-boxes with 24 XOR elements, and produces a good level of clutter. The method used in this round function is similar to methods of sorting networks. This means that the method by which cipher selects the XOR inputs is similar to methods sorting networks use to choose the (two) input elements. The key addition stage of this cipher utilizes a set of XOR elements, and its permutation layer is in the form of 4-bit element arrangements [10].

**Puffin**: This cipher was developed in 2011 by Huiju Cheng et al. It uses a 64-bit block size and a 128-bit key size, and is based on substitution-permutation network. The features of this cipher include its simple and involutive design. The SPN-based ciphers usually use several different data paths for encryption and decryption and depend on some elements to inverse the process, but the involutive nature of Puffin allows the use of encryption elements for inversion. Like many other SPN-based ciphers, the key addition stage of this cipher is implemented via an XOR operation. Its substitution layer consists of sixteen parallel 4-bit S-boxes, and its permutation layer has a bit-wise design, which if implemented in wire crossings, does not cost any hardware gates. In each round of substitution operation, cipher runs the Add-Round-key and the permutation in that order, and repeats the process for 32 round of iteration [17].

**ESF**: Eight-sided fortress was developed by LIU Xuan et al. in 2014. Like many other block ciphers, it uses 64-bit block size and 80-bit key size and is based on Feistel structure. The main component of this structure is the round function, which in this cipher is based on substitution permutation network (SPN). The aim of this cipher is to optimize the computational requirements. The round function of this cipher first subjects the 32-bit round key k and a half-block to an XOR function. The cipher then processes the output of this XOR by eight 4-bit S-boxes. The permutation layer of this round function has been designed in the form of bit permutation [6].

**Piccolo**: developed in 2011 by Kyoji Shibutani et al., this cipher uses a 64-bit block size and an 80 or 128-bit key size, and is based on type-2 Generalized Feistel Network (GFN-2). Its round function F contains eight identical S-boxes. This round function first applies four parallel 4-bit S-boxes on the input and then uses the diffusion matrix M. To produce the final output, the round function again subjects the output to four parallel 4-bit S-boxes. The permutation part of this structure is based on bit-ward permutation [14].

**Khudra**: In 2014, S. Kolay et al. developed this cipher specifically for FPGAs. This GFN2-based cipher uses a 64-bit block size and an 80 -bit key size. It utilizes two F-functions with 16-bit inputs; each F-function is based on 4-bit GFN2 structure and is employed in 6-rounds of iteration. The cipher itself uses 18 rounds of iteration. The substitution boxes used in this cipher are similar to those used in the cipher PRESENT, and have maximum algebraic degree and minimum linear-differential probability [29].

## I. EVALUATION OF LIGHTWEIGHT BLOCK CIPHERS IN TERMS OF HARDWARE-COST

In the context of hardware implementation, the term "cost" refers to the extent of occupied area; i.e. how much area does the designed hardware needs to properly operate. So obviously the ciphers with lower hardware-cost will be more desirable. Area requirement are usually measured in $\mu m^2$, but generally depend on fabrication technology and standard cell library. For independent comparison of area requirement, the more common approach is to use Gate Equivalent (GE). One GE is equal to the area required by a two-input NAND gate. So the area in GE can be obtained by dividing the occupied area in $\mu m^2$ by the area occupied by two-input NAND gate. Fig.1 shows the area occupied by lightweight block ciphers during hardware implementations.

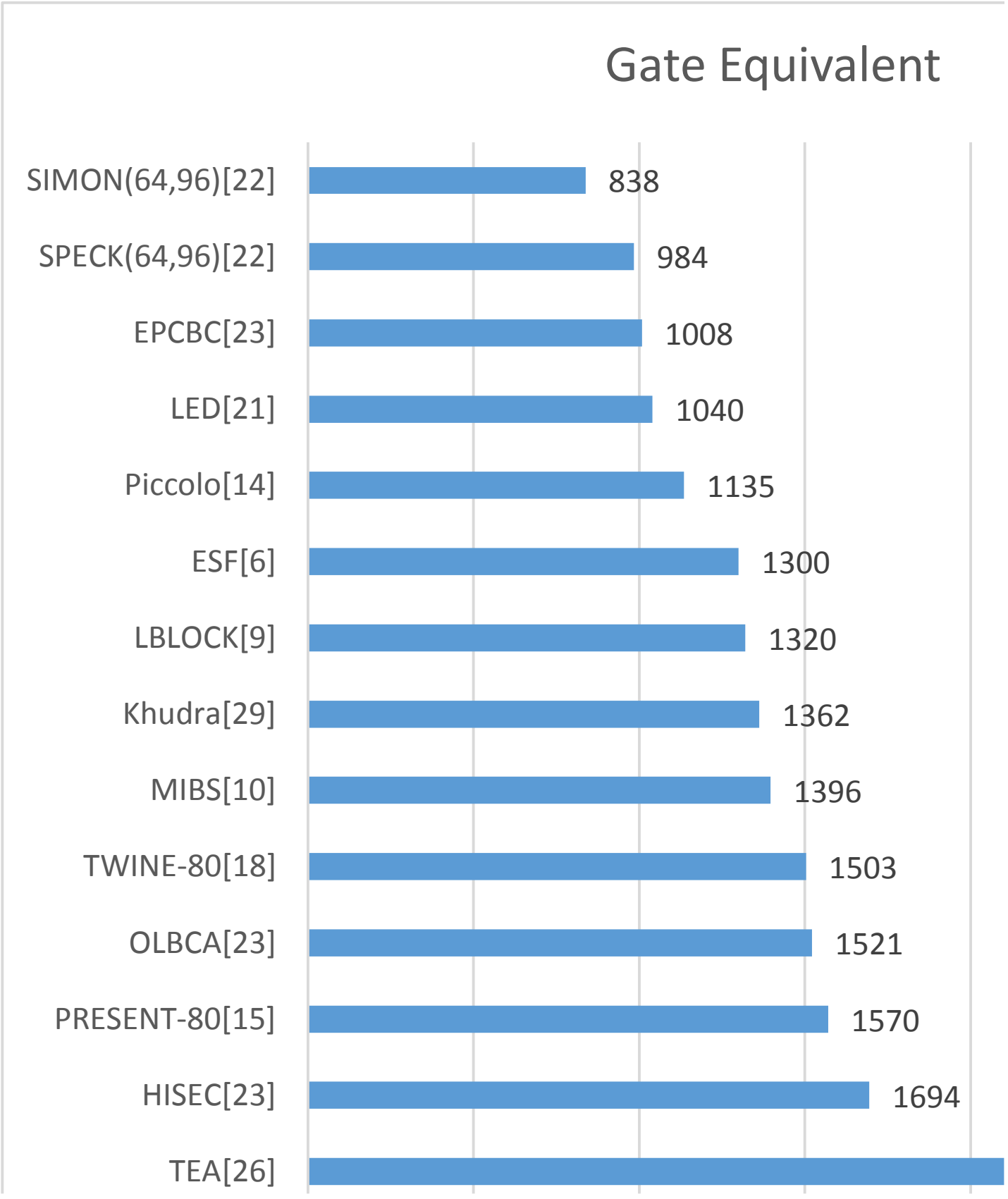

Fig.1: Evaluation of the hardware-cost of lightweight block ciphers in terms of GE

As the above chart demonstrates, from the hardware-cost perspective and with GE considered as the metric, PRINTcipher [10] shows the most desirable results, but unfortunately this cipher has a specialized and focused design and cannot provide the security required for more general applications. SEA also lacks the level of security needed for general use, as it is vulnerable to a wide range of attacks. So it seems that KTANTAN-32 is the best option in this respect. Most new ciphers use the cipher PRESENT as the basis of performance evaluation, which is due to its status as a standard cipher. The best ciphers here are SIMON and SPECK, which are relatively new, and Picollo holds the next rank. It should be iterated that from the perspective of shear GE-value, the best cipher is PRINTcipher, but it should be remembered that the best cipher is the one that achieve the best balance between, cost, security, and performance.

## II. EVALUATION OF LIGHTWEIGHT BLOCK CIPHERS IN TERMS OF SPEED

When evaluating the lightweight block ciphers in terms of their speed, the most important metrics of evaluation will be the clock cycle per block and the required time. The time required for a particular operation can be obtained by dividing the number of cycle by the operating frequency (t = cycles/freq.). To properly evaluate the lightweight ciphers with this metric, the operating frequency should be identical, so this metric fully depend on the number of cycles. This means that only one metric, either time or cycle, should be evaluated. Fig.2 shows the results of speed evaluation of lightweight block ciphers with clock cycle per block acting as the metric.

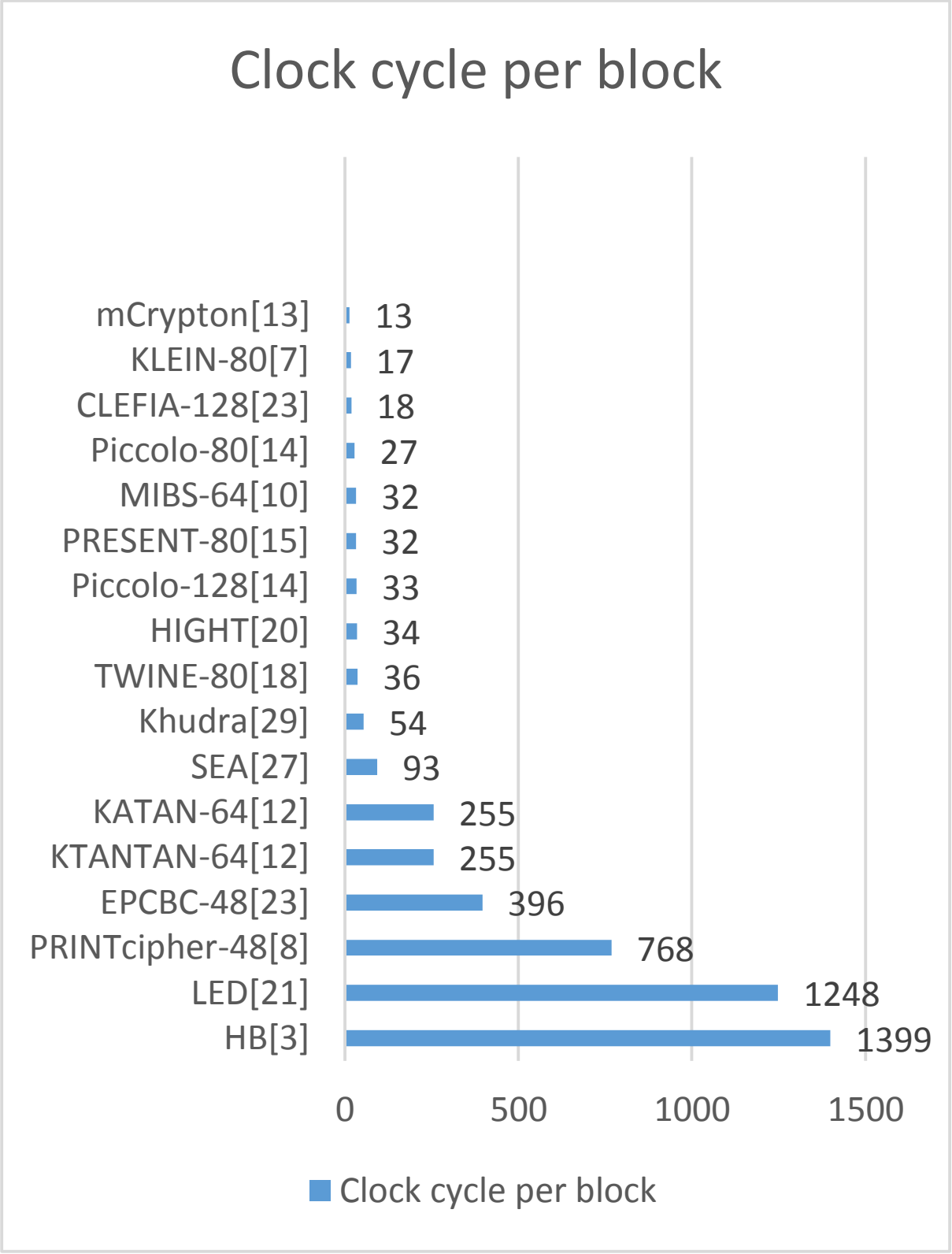

Fig.2: Evaluation of the speed of lightweight block ciphers in terms of clock cycle per block

As Fig.2 shows, from the speed perspective and with clock cycle per block acting as the metric, the best ciphers are mCrypton and KLEIN-80. The lightweight ciphers CLEFIA-128, Piccolo-80, and MIBS-64 have the next best clock cycle per block results (in that order).

## III. EVALUATION OF LIGHTWEIGHT BLOCK CIPHERS IN TERMS OF PERFORMANCE

Literature suggests that performance of lightweight ciphers can be measured by their throughput. This metric is generally defined as the rate of production, but in the context of this paper, it can be defined as the number of output bits in a specific time, or bits per second (bps). Results of evaluation of lightweight block ciphers in terms of performance with the throughput acting as the metric is presented in Fig.3.

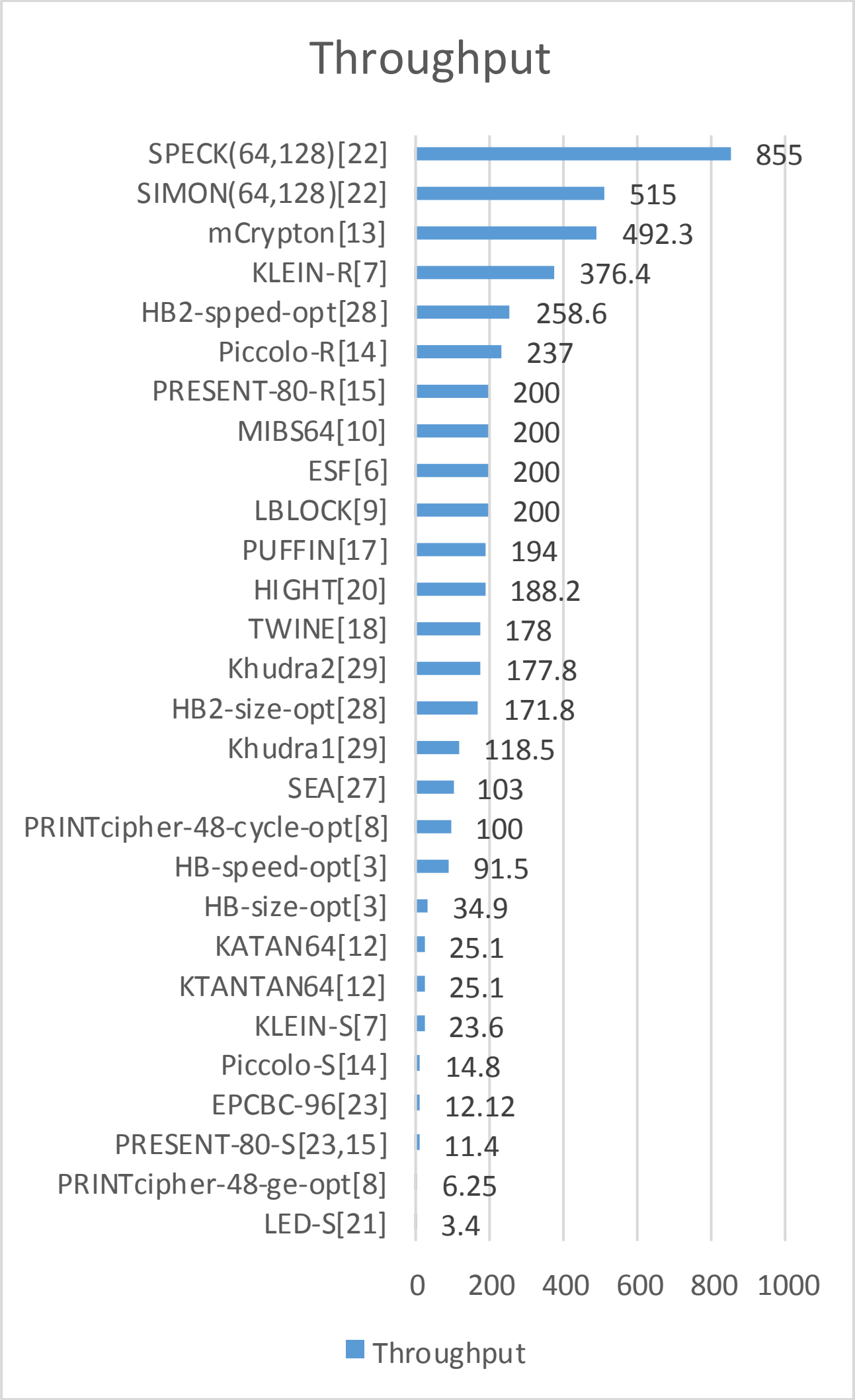

Fig.3: Evaluation of the performance of lightweight block ciphers in terms throughput (kilobits per second)

Fig.3 demonstrate that the best cipher in terms of performance (with throughput cosidered as the metric) is SPECK (64,128). The ciphers SIMON (64,128), mCrypton, KLEIN-R, and HB2-size-optimize have the next best performances in terms of throughput. Here, the symbols R and S represent round-based and serial implementations; speed-opt means cipher is implemented for the best speed, and size-opt means cipher is implemented for the best cost (or size).

## IV. Evaluation of Lightweight Block Ciphers in Terms of Balanced Efficiency

The metric used to gauge the balanced efficiency is FOM [23], which is similar to “Throughput-to-area ratio” used in [1], with only difference being the incorporation of factor $\frac{1}{GE}$. This difference stems from the fact that authors of [1] have used a uniform implementation technology, while hardware implementation conducted in [23] have been based on different technologies. The higher the FOM, the better is the cipher. The higher FOM is the result of either higher throughput or lower area squared, both pointing to a better overall efficiency. Fig.4 shows the results of hardware implementation of lightweight block ciphers with FOM considered as the metric of evaluation.

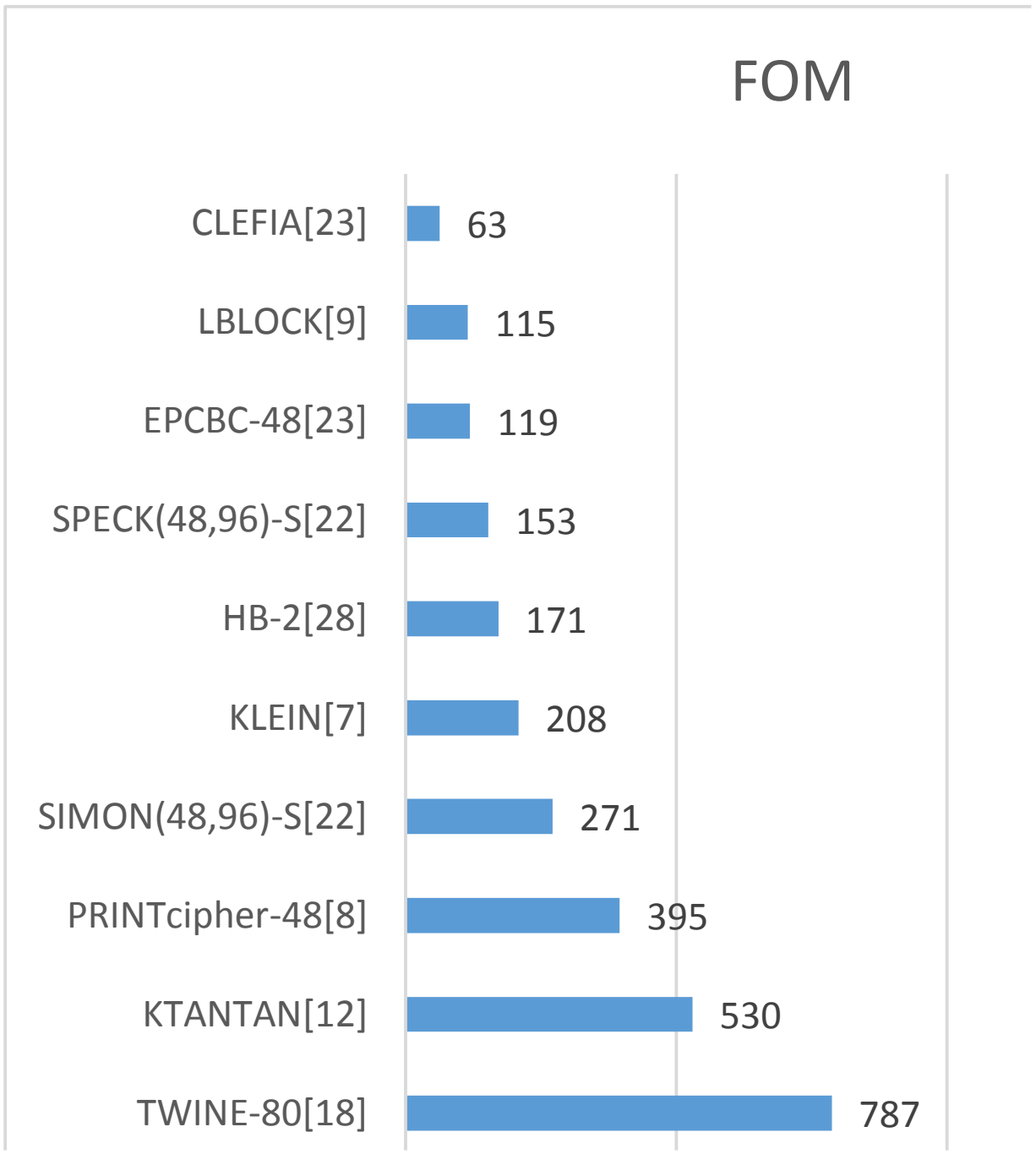

Fig.4: Evaluation of balanced efficiency of lightweight block ciphers in terms of FOM [23]

As the above figure shows, the best cipher in terms of balanced efficiency (measured by FOM) is Piccolo. The ciphers SIMON (48,96)-R and SPECK(48,96)-R have the next best performances in this respect. Here. The symbol R represents round-based implementation.

## V. Conclusion

This paper evaluated the cost, speed, performance, and balanced efficiency of lightweight block ciphers through hardware implementation. The evaluation conducted in terms of cost and with GE acting as the metric (Fig.1) determined SPECK and SIMON as the best ciphers in this respect. Piccolo held the fifth place in that category. The evaluation conducted based on clock-cycle-per-block (Fig.2) showed the better speed of mCrypton and KLEIN-80, and Piccolo was the fourth best option in this respect. The performance evaluation of lightweight block ciphers, which was based on throughput metric (Fig.3), showed the better result of SIMON and SPECK. Piccolo ranked sixth in that category. The evaluation conducted to measure the balanced efficiency of ciphers by FOM metric (Fig.4) showed that Piccolo, SIMON, and SPECK are the best ciphers in this respect. As these results show, the ciphers SIMON and SPECK exhibited the best performance with all individual metrics, and as expected, scored a decent FOM along with Piccolo. These results show that SIMON, SPECK, and Piccolo are the best lightweight block ciphers in hardware implementation.